# Dusty Molecular Cloud Collapse in the Presence of Alfvén Waves

Falceta-Gonçalves, D.; de Juli, M.C.; Jatenco-Pereira, V.

*Instituto de Astronomia, Geofísica e Ciências Atmosféricas*
*Rua do Matão, 1226, CEP 01060-970*
*Universidade de São Paulo – São Paulo – Brazil*

## Abstract

It has been shown that magnetic fields play an important role in the stability of molecular clouds, mainly perpendicularly to the field direction. However, in the parallel direction the stability is a serious problem still to be explained. Interstellar turbulence may allow the generation of Alfvén waves that propagate through the clouds in the magnetic field direction. These regions also present great amounts of dust particles which can give rise to new wave modes, or modify the pre-existing ones. The dust-cyclotron damping affects the Alfvén wave propagation near the dust-cyclotron frequency. On the other hand, the clouds present different grain sizes, which carry different charges. In this sense, a dust particle distribution has several dust-cyclotron frequencies and it will affect a broad band of wave frequencies. In this case, the energy transfer to the gas is more efficient than in the case where the ion-cyclotron damping is considered alone. This effect becomes more important if a power law spectrum is considered for the wave energy flux, since the major part of the energy is concentrated in low-frequency waves. In this work we calculate the dust-cyclotron damping in a dusty and magnetized dwarf molecular cloud, as well as determine the changes in the Alfvén wave flux. Then, we use these results to study the gravitational stability of the cloud. We show that, considering the presence of charged dust particles, the wave flux is rapidly damped due to dust-cyclotron damping. Then the wave pressure acts in a small length scale, and cannot explain the observable cloud sizes, but can explain the existence of small and dense cores.

# 1 – Introduction

There are in the literature several works concerning the mechanical stability of the dwarf molecular clouds (DMC). These clouds have masses $< 10^3$ $M_\odot$ in regions of 2 – 5 $pc$ of radius, and temperature of ~ 10 – 20 $K$ (Shu, Adams & Lizano 1987, Evans II 1999). It is believed that these objects can live more than $10^8$ years in equilibrium. A cloud with only thermal support will collapse if the mass exceeds the Jeans mass. This fact seems to be a problem in DMC considering that the Jeans mass of an isothermal gas is

$$M_J \equiv \rho_0 \lambda_J^3 = \left(\frac{\pi}{G}\right)^{3/2} \rho_0^{-1/2} c_s^3, \qquad (1)$$

where $\rho_0$ is the mass density and $c_s$ is the sound velocity. In the case of DMCs, $M_J$ is a few solar masses, which is much smaller than the cloud mass. Similarly, the free-fall collapse time,

$$t_{ff} = \left(\frac{3\pi}{32 G \rho_0}\right)^{1/2}, \qquad (2)$$

is ~ $10^6$ years, which is much smaller than the lifetime of these objects. As a consequence, the thermal pressure alone cannot explain the stability of these clouds. Several additional support mechanisms have been proposed, like magnetic fields (e.g. Chandrasekhar & Fermi 1953), rotation (Field 1978) and turbulence (Norman & Silk 1980; Bonazzola *et al.* 1987).

Typical magnetic fields of ~ µGauss (Crutcher 1999) can increase $M_J$ to the expected values. However, the magnetic pressure acts against the collapse only in the perpendicular direction of the magnetic field. Turbulence can excite the generation of MHD waves, which can propagate along the field lines, adding an extra pressure term in the parallel direction (McKee & Zweibel 1995; Gammie & Ostriker 1996; Martin, Heyvaerts & Priest 1997). Martin *et al.* (1997) showed that an Alfvén wave flux can support the cloud collapse in the parallel direction to the magnetic field, but for that, it would be necessary a great amount of magnetic energy concentrated in a weakly damped wave spectrum range.

There are in the literature several damping mechanisms for the Alfvén waves. In particular, Martin *et al.* (1997), studying the changes on the DMCs stability due to the presence of Alfvén waves propagating along the magnetic field, considered the damping of these waves due to the ion-neutral collisions, which is weak for low frequency waves. Zweibel & Josafatsson (1983) also showed that if only ion-neutral collisional and non-linear damping mechanisms are considered, the Alfvén wave mode is weakly damped in DMC. In this case, the wave flux may reach the edges of the cloud and support the gas against gravity. However, if strong damping mechanisms may take place, the wave flux will be damped suddenly, and will not reach the edges of the cloud.

The physical conditions of DMC suggest the existence of charged dust particles due to collisions with electrons of the plasma. Chhajlani & Parihar (1994) showed that charged dust particles affect the gravitational stability, changing the Jeans criterion both on perpendicular and parallel directions to the field. In another view, once charged, these particles will suffer the influence of the magnetic field that gives rise to a cyclotron frequency and a resonance associated to them. For the ions, this resonance occurs in a narrow range of higher frequencies, being unimportant in the systems under consideration. On the other hand, the dust cyclotron resonance occurs in low frequencies, and it can be an important damping mechanism for the waves propagating in molecular clouds. Mathis, Rumple & Nordsiek (1977) (MRN) observed and fitted the interstellar medium (ISM) extinction to a distribution of different dust particles. In particular, the dust particles size distribution seems to be a power law spectrum, $f(a) = C \cdot a^{-p}$, with $p \sim 3 - 4$, for different dust compositions. Tripathi & Sharma (1996) and Cramer, Verheest & Vladimirov (2002) showed that a size distribution of charged dust particles can modify the dispersion relation of Alfvén waves and give rise to a wave resonant damping in the frequencies coincident to the dust-cyclotron frequency.

In this work we present a model in which a flux of waves propagating in a dwarf molecular cloud is damped due to resonant interaction with dust charged particles, and we analyze the consequences for the cloud stability. We proceed as follows. In section 2, we describe the wave dispersion relation modified by the presence of charged dust. In section 3, we describe the model for the cloud stability, and present the results. Finally, we draw the conclusions of the paper.

## 2 – Alfvén waves propagating in a dusty plasma

The propagation and damping of Alfvén waves in dusty plasma has been considered by many authors (e.g. Pillip, Morfill, Hartquist & Havnes 1987; Mendis & Rosenberg 1992; Shukla 1992). Although the number of the dust particles is smaller than the ions one, the process of dust charging is efficient and these particles can obtain charges in the order of $q_d \sim 10^0 - 10^3$ $e^-$ in astrophysical media (Goertz 1989; Mendis & Rosenberg 1994). Dust particles modify the plasma behavior in different ways. In particular, charged dust particles introduce a cutoff in the Alfvén wave in the dust cyclotron frequency. If a distribution of grain sizes is considered, we obtain a large band of resonance frequencies instead of a single one.

According to MRN, we can describe the distribution of grain sizes by the function $f(a) = C \cdot a^{-p}$ where $a$ is a dimensionless radius, defined as $a = r / r_{\min}$, where $r_{mín} < r < r_{máx}$, and $C = (p-1)/(1-a_m^{1-p})$. We define $a_m \equiv r_{\max}/r_{\min}$ as the ratio of maximum and minimum dust radii. Observationally, the parameter $p$ depends on the dust constituent and on the environment. Typically, for dust particles of $0.005 \mu m < r < 1 \mu m$, we have $3 \leq p \leq 4$. In this work we will consider $p = 4$, as used by Cramer *et al*. (2002), and, in this case, $C = 3/(1-a_m^{-3})$.

The wave propagation in dusty plasma is modified, and some new and interesting effects take place (Wardle & Ng 1999; Cramer 2001). The dispersion relation of the Alfvén waves, with frequencies smaller than the ion cyclotron frequency, considering constant

charged dust particles in a neutral and cold dusty plasma, is given by (Cramer *et al.* 2002):

$$k_z^2 = u_1 \pm u_2, \tag{3}$$

where

$$u_1 = \frac{\omega^2 \Omega_{i0}^2}{v_A^2 (\Omega_{i0}^2 - \omega^2)} + \frac{\omega^2 \Omega_{d\max}^2}{sv_{Ad}^2} \int_1^{a_m} \frac{f(a)}{a\left(\left(\Omega_{d\max}^2/a^4\right) - \omega^2\right)} da, \tag{4}$$

and

$$u_2 = \frac{\omega^3 \Omega_{i0}}{v_A^2 (\Omega_{i0}^2 - \omega^2)} + \frac{\omega^3 \Omega_{d\max}}{sv_{Ad}^2} \int_1^{a_m} \frac{f(a)}{\left(\left(\Omega_{d\max}^2/a^4\right) - \omega^2\right)} da, \tag{5}$$

where $s = C \cdot \ln(a_m)$, $B_0$ is the external mean magnetic field, $\omega$ is the angular wave frequency, $v_A = B_0/\sqrt{4\pi\rho_i}$ is the Alfvén speed in terms of the ion density, $v_{Ad} = B_0/\sqrt{4\pi\rho_d}$ is the Alfvén speed, $\rho_d$ is the dust mass density, $\Omega_{i0} = q_i B_0/m_i c$ is the ion cyclotron frequency and $\Omega_{d\max} = q_d B_0/m_d c$ is the maximum dust cyclotron frequency (*i.e.* for the minimum dust radius). The mean grain particle charge can be obtained for each dust radius considering charging current equilibrium over dust surface. The equilibrium equation, considering Maxwellian distribution of velocities, is given by:

$$\frac{\omega_{pi}^2}{v_{Ti}} \left(1 + \frac{q_d q_e}{rk_B T}\right) = \frac{\omega_{pe}^2}{v_{Te}} e^{-q_d q_e/rk_B T}, \tag{6}$$

where $\omega_{p\beta}$ is the plasma frequency and $v_{T\beta}$ is the thermal velocity of the $\beta$ specie.

The left-hand polarized wave (-) interacts with ions, and is not affected by the negatively charged dust particles resonance. The right-hand polarized wave (+) is the mode damped by the dust resonance. In the case of $\Omega_{d\max}/a_m^2 < \omega < \Omega_{d\max}$, the integral in the particles radii has singularities, whose residues give the complex part of the wave number ($k_i$) and that leads to the dust-cyclotron damping of the waves. If dust density vanishes (*i.e.* considering a dustless plasma), $v_{Ai}/v_{Ad} \propto (\rho_d/\rho_i)^{1/2} = 0$. By that, multiplying equations (3), (4) and (5) by the ion Alfvén velocity ($v_A$) the term that gives the imaginary part of the wave number will vanishes, and equation (3) resumes to the well known dispersion relation for ion and electrons plasma.

Cramer *et al.* (2002) solved equations (3) – (5) using $a_m = 1.1$. However, in the interstellar gas, this ratio is ~ $10^2$. We performed same calculations, including the determination of mean dust particle charge, using equations (3) – (6). To illustrate how this parameter influences the wave dispersion relation, we plotted in Figure 1 the real and imaginary parts of the wave number, using different values for the $a_m$ parameter.

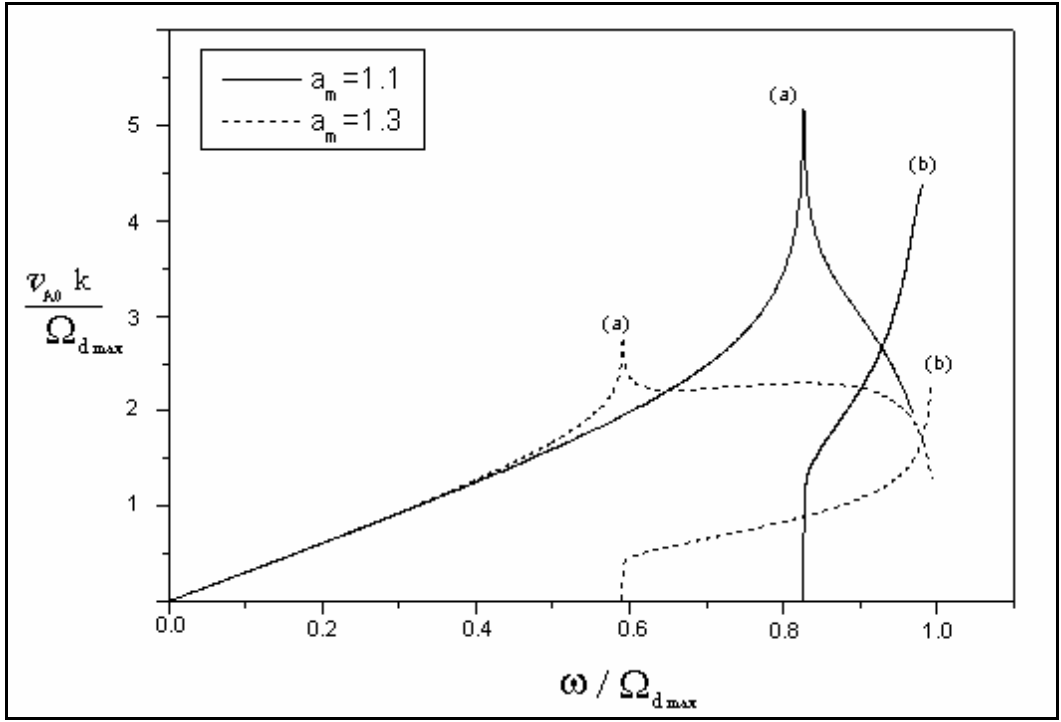

Figure 1: (a) The real part of the wave number for $a_m = 1.1$ (solid line) and $a_m = 1.3$ (dotted line). (b) The imaginary part of the wave number for $a_m = 1.1$ (solid line) and $a_m = 1.3$ (dotted line).

In Figure 1, we plotted the solution of the equation (3) using as a parameter the dust mass density, given by $\rho_{H_2}/\rho_d \sim 200$ (Kramer *et al*. 2003; Perna, Lazzati & Fiore 2003; Spitzer 1968). In (a) we have the real part of the wave number for $a_m = 1.1$ (solid line), used also by Cramer *et al*. (2002), and $a_m = 1.3$ (dotted line), chosen to illustrate how a small change in $a_m$ induces an important modification in the dispersion relation. In (b) we have the imaginary part of the wave number for $a_m = 1.1$ (solid line), and $a_m = 1.3$ (dotted line). For $a_m = 1.1$ the resonance band occurs in the range $0.83 < \omega/\Omega_{d\,max} < 1$, and in the case of $a_m = 1.3$ the resonance band occurs for $0.59 < \omega/\Omega_{d\,max} < 1$. Note that the range of resonant frequencies is then proportional to $a_m^2$. In the interstellar medium we have $a_m > 100$ and the expected resonance band would range from $0.0001 < \omega/\Omega_{d\,max} < 1$, affecting almost all low frequency spectrum. By this fact, it is interesting to study the effects of the dust cyclotron damping on the Alfvén wave propagation, and the consequences in the mechanical stability of DMC´s due to the presence of these particles.

## 3 – The cloud stability

Following the model proposed by Martin *et al*. (1997), we will study the case of a flattened cloud, permeated by an uniform magnetic field $\vec{B} = B_0\hat{z} + \delta B\hat{x}$, whose perturbation $\delta B$ is an Alfvén wave propagating ($\vec{k}$) in the parallel direction to the magnetic field. The wave pressure allows another cloud support mechanism, as shown in the scheme on Figure 2. We considered that $\delta B << B_0$ so that the linear approach can be used.

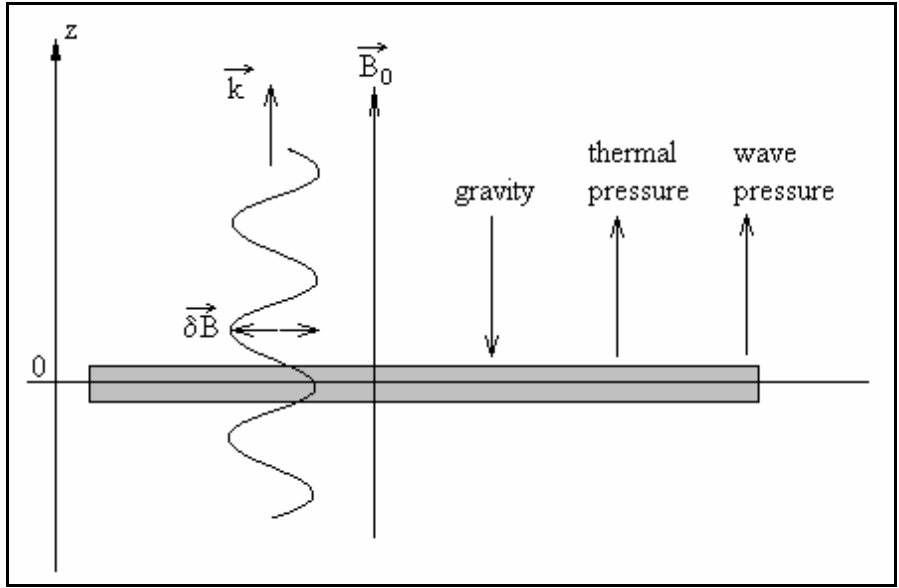

Figure 2: In this scheme, we show the propagation direction of the Alfvén waves and the way as they increase the cloud support against gravity.

The momentum equation of a self-graviting plasma, including magnetic pressure, is given by:

$$\rho\frac{\partial \vec{u}}{\partial t} + \rho(\vec{u}.\vec{\nabla})\vec{u} = -\vec{\nabla}P + (\vec{B}.\vec{\nabla})\frac{\vec{B}}{4\pi} - \vec{\nabla}\left(\frac{B^2}{8\pi}\right) + \rho\vec{g}, \qquad (7)$$

where, $u$ is the velocity, $P$ is the thermal pressure of the gas, $\rho$ is the mass density, $B$ is the magnetic field and $g$ is the gravity acceleration.

In the perpendicular direction to $\vec{B}_0$, the magnetic field pressure is sufficient to prevent the gravitational collapse. For this reason, we will study the stability in the *z*-direction. Considering the model described in Figure 2, the temporal mean of the momentum equation (7) for a cloud in mechanical equilibrium can be written by:

$$-\vec{\nabla}P - \vec{\nabla}\varepsilon + \rho\vec{g} = 0, \qquad (8)$$

where $\varepsilon = <\delta B^2>/8\pi$ is the magnetic energy density of the MHD waves propagating in the parallel direction to $\vec{B}_0$.

As we will see in the following section, the waves that are affected have frequencies near the dust-cyclotron frequency, which leads to wavelengths of $\sim 10^{15}$ cm. In this case, since the waves have a wavelength smaller than the system scales ($\sim 0.1$ pc), the WKB approach can be used and the energy conservation equation will be given by:

$$\vec{\nabla} \cdot \ln(\varepsilon \cdot v_A) = -L^{-1}, \tag{9}$$

where $L$ is the wave damping length, which is related to the imaginary part of the wave number ($k_i$) determined in the previous section, by $L = 2\pi / k_i$.

The system of differential equations is completed with the Poisson equation

$$\vec{\nabla} \cdot \vec{g} = 4\pi G \rho, \tag{10}$$

where $G$ is the gravitational constant.

Tu, Roberts & Goldstein (1989) observed an Alfvén wave spectrum propagating in the solar wind, and inferred the following power law spectrum,

$$\Phi_A(\omega) = \Phi_A(\omega_0) \cdot \left(\frac{\omega}{\omega_0}\right)^{-\alpha}, \tag{11}$$

with $\alpha \sim 0.6$ and $\Phi_A(\omega_0)$ is the Alfvén wave flux at the frequency $\omega_0$.

The origin of the Alfvén wave spectrum is unknown, and many possibilities have been proposed to explain it, *e.g.* turbulence, magnetic field annihilation and convection. We will suppose that some internal process in DMC´s generates magnetic perturbations, which act as the origin of a spectrum of the Alfvén waves similar to that observed in the solar wind. The total flux is given by:

$$\Phi_{tot} = \int_{\omega_{min}}^{\omega_{max}} \Phi_A(\omega) d\omega. \tag{12}$$

The relation between the total flux and the wave energy density is $\varepsilon = \Phi_{tot} / v_A$, and it is related to the thermal energy density of the gas ($U_{int}$) by the parameter $\Lambda = \varepsilon / U_{int}$. Thus, for a particular choose of the free parameter, $\Phi_A(\omega_0)$, we calculate the total flux and the wave energy density, which is related to the internal energy density by $\Lambda$.

In the following sub-sections we will describe the cloud stability, by the solutions of equations (8) – (10), considering two cases: (i) when there is no damping of the Alfvén waves, and (ii) when the dust-cyclotron damping is considered for different $\Lambda$ values. Low $\Lambda$ values guarantee that the linear approach can be used. Typically, the gas pressure to the magnetic pressure ratio: $\beta = \dfrac{P_g}{B_0^2 / 8\pi}$, assumes low values for molecular clouds ($\beta \sim 0.04$)

(Crutcher 1999; Sigalotti & Klapp 2000), implying that magnetic fields are extremely important in these regions. If one assumes a high flux of Alfvén waves, i.e. $\Lambda > 3$ (as done by Martin *et al.* 1997), the Alfvén wave amplitude to the mean magnetic field ratio ($\eta = \delta B / B_0$) will be greater than 0.1, and in this case the linear approximation is no more valid.

*3.1 – Cloud Stability: Alfvenic support without damping*

If there is no wave damping, $L^{-1} \to 0$, the wave energy density ($\varepsilon$), given by equation (9), reduces to $\varepsilon(z) = \varepsilon(z=0)\left(\rho/\rho_0\right)^{1/2}$. In this case, a gradient in density produces a gradient in $\varepsilon$, and the latter acts as a support against the collapse of the cloud. Substituting this expression in equation (8), and using also equation (10), the density profile solution can be obtained. The result is shown in Figure 3. The local parameters used are: the central number density of the dwarf molecular cloud $n_0 = 10^4 cm^{-3}$, temperature $T = 20K$ and mean magnetic field $B_0 = 10\mu G$ (Evans II 1999). The solid line shows the density profile in the case without Alfvenic support ($\Lambda = 0$), and the other plots represent equilibrium with Alfvenic support for three different $\Lambda$ values: $\Lambda = 0.05$, $\Lambda = 0.15$ and $\Lambda = 0.25$.

Figure 3 shows that the equilibrium cloud size increases as the energy density of the waves ($\varepsilon$) propagating through the cloud increases, in accordance to Martin *et al.* (1997). This indicates that the observed properties of these objects could be explained by an Alfvén waves flux. However, if we consider the wave damping, this result is modified, as we show in the next sub-section.

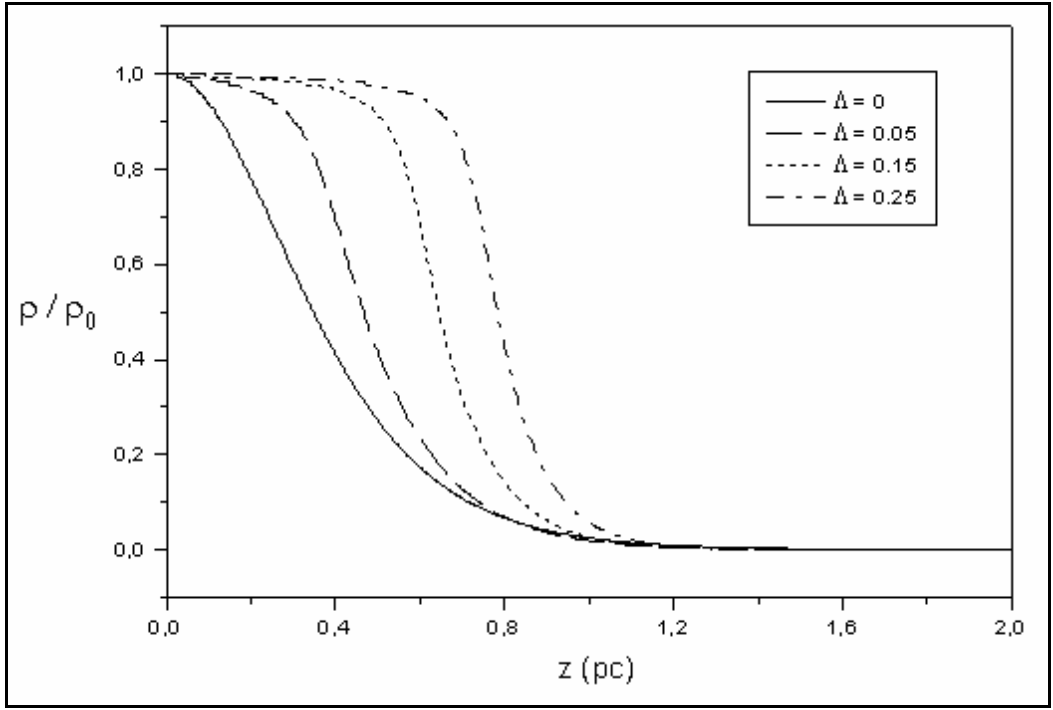

Figure 3: The cloud density profile as a function of distance for different values of the parameter $\Lambda$ for waves without damping. The solid line represents the situation without Alfvenic support, and the dashed, dotted and dot-dashed lines represent the cases for $\Lambda = 0.05$, $\Lambda = 0.15$ and $\Lambda = 0.25$, respectively.

*3.2 – Cloud Stability: Alfvenic support with damping*

In a more realistic model, it is necessary to consider the wave propagation in a dusty medium, whose dust particles are charged. In this case, the dust-cyclotron damping treated in Section 2 takes place. The calculation of equation (6) gives, for the cloud parameters used in the previous section, a mean dust charge of $\bar{q}_d \cong -1e^-$. Using the dust-cyclotron damping already described, and introducing equation (11) into equation (9), the modifications in the energy flux spectrum along the cloud *z*-direction can be obtained. The damping length, in function of the wave frequency, is given in Figure 4. The Alfvén wave spectrum, modified as the waves propagate through the cloud, is shown in Figure 5.

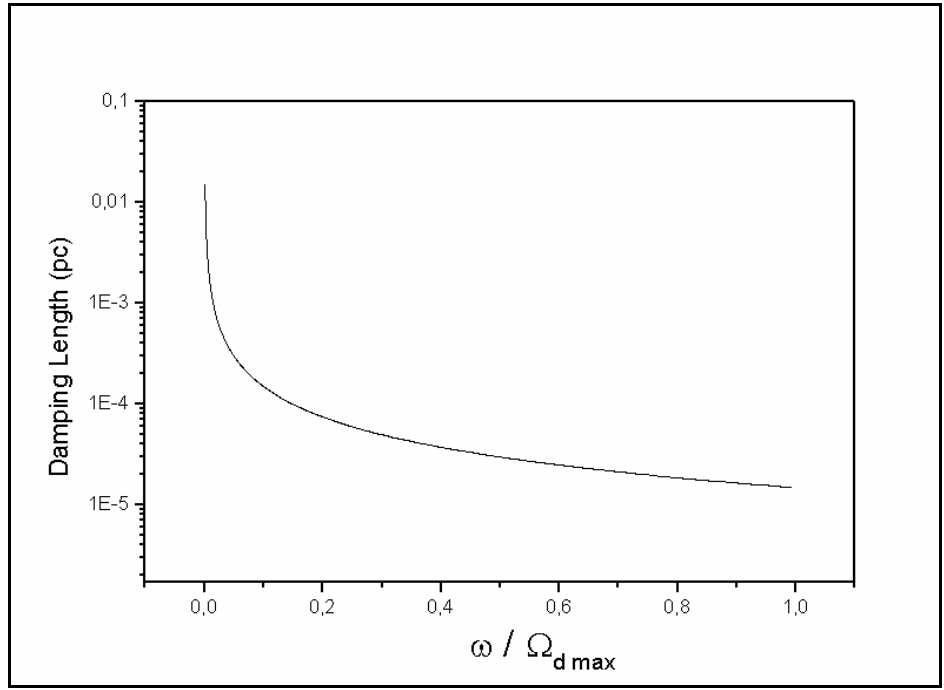

Figure 4: The wave dust cyclotron damping length as function of wave frequency for the chosen molecular cloud parameters.

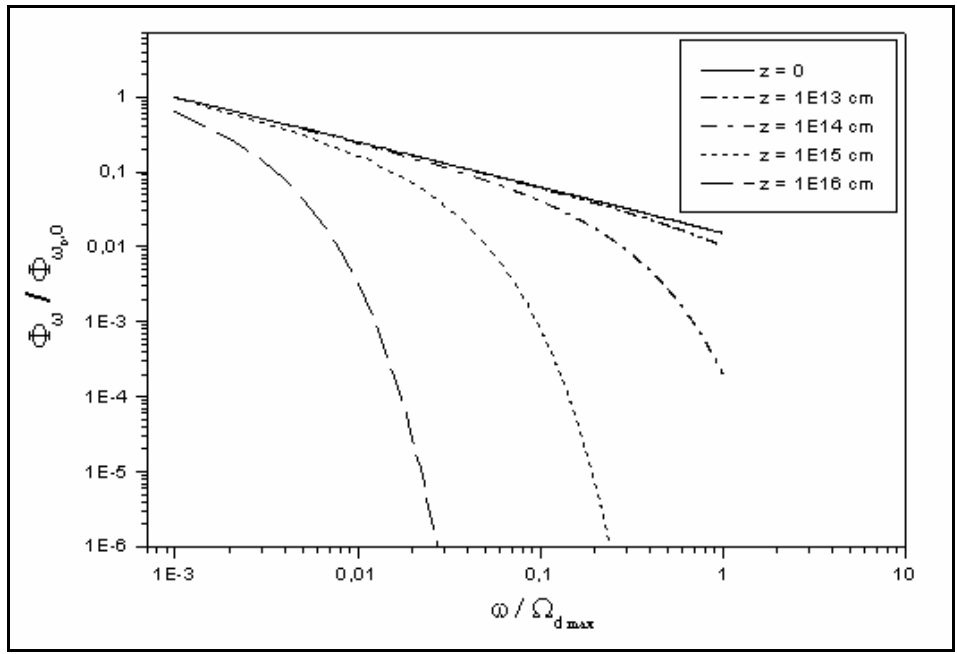

Figure 5: The wave power spectrum, damped by dust-cyclotron resonance, for different locations ($z$) in the cloud.

In Figure 5 we notice that the waves flux, given by equation (11), is damped in the frequency range of dust-cyclotron resonance. The waves with frequencies coincident to the dust-cyclotron frequencies are damped ($\omega \sim \Omega_{d\,max}$). We also note that the frequency band is almost completely damped up to $z \sim 10^{-2}$ pc. In our calculations we used in equation (11), $\alpha = 0.6$. The choice of this parameter does not change our conclusions since it just represents how energy is concentrated in frequency spectrum. If α is too large, a few energy is carried by high frequency waves and most of energy would be carried by low frequency waves, which continues to be damped within $z \sim 10^{-2}$ pc.

The cloud density profile, considering wave damping, can be obtained by the solution of the equations (8) – (12), and it is shown in Figure 6. The dotted line shows the density profile in the case without Alfvenic support ($\Lambda = 0$), and the others represent equilibrium with Alfvenic support including the damping mechanism for three different $\Lambda$ values: $\Lambda = 0.05$, $\Lambda = 0.15$ and $\Lambda = 0.25$.

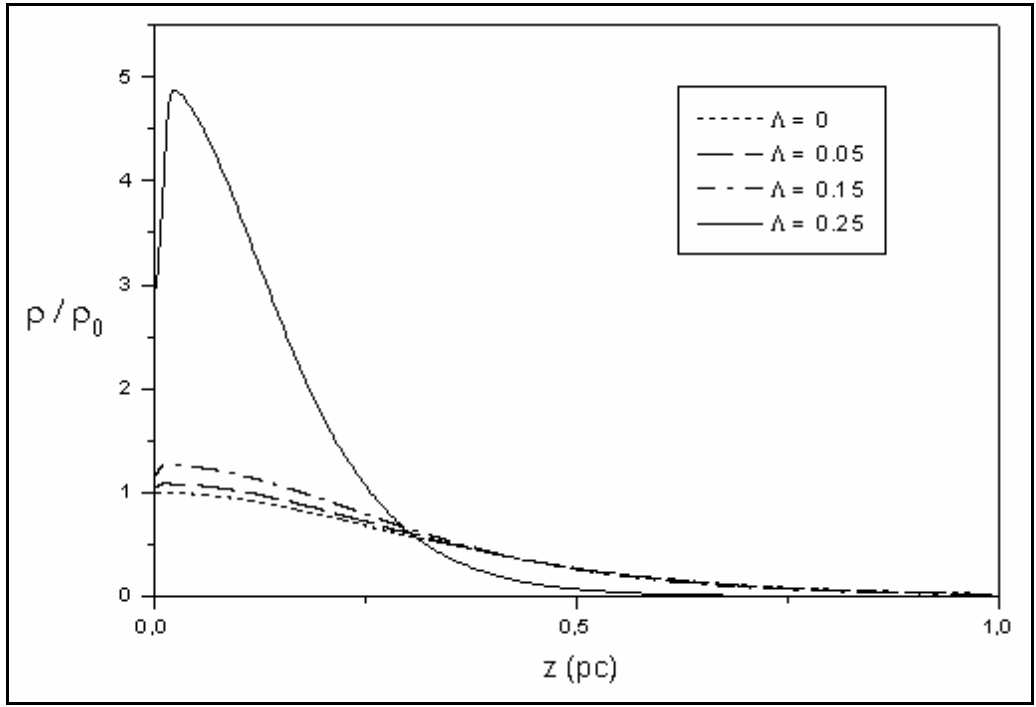

Figure 6: The cloud density profile as a function of distance for different values of the parameter $\Lambda$, including the wave damping. Dotted line represents equilibrium without Alfvenic support, and the dashed, dot-dashed and solid lines represent the cases of $\Lambda = 0.05$, $\Lambda = 0.15$ and $\Lambda = 0.25$, respectively.

As we can see in Figure 6, the waves are damped for $z < 1\,pc$ and cannot support the edges of molecular clouds. This result can be compared to Figure 3 of Martin *et al.* (1997), where a weak Alfvén wave damping allows the stability of the cloud up to several parsecs. In this sense, the damped Alfvenic support can not be used to explain the observed cloud sizes of $z \sim 1-5\,pc$, unless one assumes that the waves are being generated along all the cloud. However, it is important to notice that the sudden damping of the wave flux results in compact and denser cloud cores. In this case, dense and compact cores, or even cloud clumpiness, already observed in DMC´s, *e.g.* by Snell *et al.* 1984 and Loren, Sandqvist & Wooten 1983, could be explained.

**Conclusions**

Considering typical dwarf molecular clouds with number density $n_{H_2} \sim 10^4 \, cm^{-3}$ and temperature T ~ 20*K*, the corresponding Jeans mass is $M_J$ ~ 3 $M_\odot$. If this result is compared to the observed cloud masses $M \leq 100 M_\odot$, we notice that it is necessary an additional mechanism acting together with thermal pressure to support the cloud against its own gravity. Magnetic field of ~ µGauss has been observed in these regions, which can yield an extra pressure term to keep the stability perpendicularly to the field lines. However, for the direction parallel to the magnetic field lines, where there is no pressure from the mean magnetic field, it is necessary other mechanism to support the cloud. Turbulence can excite MHD waves, in particular, the uncompressive Alfvén wave mode that can propagate along the magnetic field lines. It is believed that these waves are slowly damped in molecular clouds. In this case, we show that they can add a wave pressure term, and support these clouds.

In a more realistic model, we use the fact that these clouds present also a great amount of charged dust particles that suffer the influence of the magnetic field. These particles have a gyro-frequency that can give rise to a dust-cyclotron resonance with the Alfvén waves frequency (Tripathi & Sharma 1996; Cramer *et al.* 2002). In this interaction the wave is damped and a gradient in the wave flux is established. In this work we have shown that, when wave damping is not considered, the wave flux can support the cloud against gravity, preventing its collapse, as also pointed by Martin *et al.* (1997). On the other hand, considering the existence of charged dust particles, the waves are strongly damped due to the dust cyclotron damping. Taking into account this wave damping, discussed by Cramer *et al.* (2002), the flux is dissipated suddenly (in a region << 1*pc*) leading to the formation of a compact and dense core. In this case, the waves could not reach the out layers of the cloud, and if so, they could not be used to explain the size of these objects, although they still could be used to inhibit star formation.

**Acknowledgements**

The authors would like to thank the Brazilian agencies CNPq and FAPESP (No. 01/10921-9) for financial support. The authors would also like to thank the project PRONEX/FINEP (No. 41.96.0908.00) for partial support.

## References

- Bonazzola, S. *et al.*: 1987, A&A, **172**, 293.
- Chandrasekhar, S. & Fermi, E.: 1953, ApJ, **118**, 116.
- Chhajlani, R. & Parihar, A.: 1994, ApJ, **422**, 746.
- Cramer, N., Verheest, F. & Vladimirov, S.: 2002, Phys. Plasmas, **9**, 4845.
- Cramer, N.: 2001, in "The Physics of Alfvén Waves", Wiley, Berlin.
- Crutcher, R.: 1999, ApJ, **520**, 706.
- Evans II, N.: 1999, ARAA, **37**, 311.
- Field, G.: 1978, in "Protostars & Planets", University of Arizona Press, 243.
- Gammie, C. & Ostriker, E.: 1996, ApJ, **466**, 814.
- Goertz, C.: 1989, Rev. Geophys., **27**, 271.
- Kramer, C., Richer, J., Mookerjea, B., Alves, J. & Lada, C.: 2003, A&A, **399**, 1073.
- Loren, R., Sandqvist, A. & Wooten, A.: 1983, ApJ, **270**, 620.
- McKee, C. & Zweibel, E.: 1995, ApJ, **440**, 686.
- Martin, C., Heyvaerts, J. & Priest, E.: 1997, A&A, **326**, 1176.
- Mathis, J., Rumpl, W. & Nordsiek, K.: 1977, ApJ, **217**, 425. (MRN)
- Mendis, D. & Rosenberg, M.: 1992, IEEE Trans. Plasma. Sci., **20**, 929.
- Mendis, D. & Rosenberg, M.: 1994, ARAA, **32**, 419.
- Norman, C. & Silk, J.: 1980, ApJ, **238**, 158.
- Perna, R., Lazzati, D. & Fiore, F.: 2003, ApJ, **585**, 775.
- Pillip, W., Morfill, G., Hartquist, T. & Havnes, O.: 1987, ApJ, **314**, 341.
- Shu, F., Adams, F. & Lizano, S.: 1987, ARAA, **25**, 23.
- Shukla, P.: 1992, Phys. Scr., **45**, 504.
- Sigalotti, L. & Klapp, J.: 2000, ApJ, **531**, 1037.
- Snell, R., Goldsmith, P., Erickson, N. *et al.*: 1984, ApJ, **276**, 625.
- Spitzer, L.: 1968, "Diffuse Matter in Space", InterScience Publishers, New York.
- Tripathi, K. & Sharma, S.: 1996, Phys. Plasmas, **3**, 4380.
- Tu, C., Roberts, D. & Goldstein, M.: 1989, JGR, **94**, 13575.
- Wardle, M. & Ng, C.: 1999, MNRAS, **303**, 239.
- Zweibel, E. & Josafatsson, K.: 1983, ApJ, **270**, 511.